\documentclass[useAMS,usenatbib]{mn2e}
\newcommand{\cri}{\textsc{C.i}}
\newcommand{\crii}{\textsc{C.ii}}
\newcommand{\criii}{\textsc{C.iii}}

\usepackage{graphicx}
\usepackage{amssymb,amsmath}
\usepackage{xcolor}
\usepackage{url}    %base package for URL support
\usepackage[
      colorlinks=false,    %no frame around URL
      urlcolor=black,    %no colors
      citecolor=black,
      urlcolor=black,
      menucolor=black,    %no colors
      linkcolor=black,    %no colors
      pagecolor=black,    %no colors
      bookmarks=true,    %tree-like TOC
      citebordercolor=lightgray,
      linkbordercolor=lightgray,
      urlbordercolor=lightgray,
      bookmarksopen=true,    %expanded when starting
      hyperfootnotes=false,    %no referencing of footnotes, does not compile
      pdfpagemode=UseOutlines    %show the bookmarks when starting the pdf viewer
]{hyperref}

\title[Spectral-timing of SWIFT~J1753.5--0127]{Accretion flow diagnostics with X-ray spectral-timing: the hard state of SWIFT~J1753.5--0127}

\author[P. Cassatella, P. Uttley \& T.~J. Maccarone]{P. Cassatella$^{1}$\thanks{E-mail:
Pablo.Cassatella@soton.ac.uk}, P. Uttley$^{2}$ and T.~J.
Maccarone$^{1}$\\
$^{1}$Astronomy Group, Faculty of Physical and Applied Sciences,
University of Southampton, Southampton SO17 1BJ, United Kingdom\\
$^{2}$Astronomical Institute `Anton Pannekoek', University of Amsterdam,
P.O. Box 94249, 1090 GE Amsterdam, the Netherlands
}

\begin{document}

\date{}
\pagerange{\pageref{firstpage}--\pageref{lastpage}} \pubyear{2012}

\maketitle

\label{firstpage}

\begin{abstract}
Recent {\it XMM-Newton} studies of X-ray variability in the hard
states of black hole X-ray binaries (BHXRBs) imply that the variability is generated in the `standard' optically-thick accretion disc that is responsible for the multicolour black-body emission. The variability originates in the disc as mass-accretion fluctuations and propagates through the disc to `light up' inner disc regions, eventually
modulating the power-law emission that is produced relatively centrally.  Both the covariance spectra and time lags that cover the soft
band strongly support this scenario. 

Here we present a comparative spectral-timing study of {\it XMM-Newton} data from the BHXRB
SWIFT~J1753.5--0127 in a bright 2009 hard state with that from the
significantly fainter 2006 hard state, to show for the first time the
change in disc spectral-timing properties associated with a global increase in both the accretion rate and the relative contribution of the disc emission to the bolometric luminosity.

We show that, although there is strong evidence for intrinsic disc
variability in the more luminous hard state, the disc variability amplitude is suppressed relative to that of the power-law emission, which contrasts with the behaviour at lower luminosities where the disc variability is slightly enhanced when compared with the power-law variations.  Furthermore, in the higher-luminosity data the disc variability below 0.6~keV becomes incoherent with the power-law and higher-energy disc emission at frequencies below 0.5~Hz, in contrast with the coherent variations seen in the 2006 data.  We explain these differences and the associated complex lags in the 2009 data in terms of the fluctuating disc model, where the increase in accretion rate seen in 2009 leads to more pronounced and extended disc emission.  If the variable signals are generated at small radii in the disc, the variability of disc emission can be naturally suppressed by the fraction of unmodulated disc emission arising from larger radii.  Furthermore the drop in coherence can be produced by disc accretion fluctuations arising at larger radii which are viscously damped and hence unable to propagate to the inner, power-law emitting region.
\end{abstract}


%%%%%%%%%%%%%%%%%%%%%%%%%%%%%%%%%%
\section{Introduction}
%%%%%%%%%%%%%%%%%%%%%%%%%%%%%%%%%%
\label{sec:intro}

Variability on a broad range of time-scales, ranging from milliseconds to
hours, is a well-known characteristic of black hole X-ray binaries (BHXRBs).
The short-term variability (time-scales up to a few hundred seconds) is
strongly dependent on the spectral state of the object, being
particularly enhanced in the so-called hard state, with rms
amplitudes up to $\sim$ 40 per cent of the average flux
\citep{belloni2005,remillard2006,munozdarias2011}.

In the hard state, the X-ray emission is dominated by a hard power-law
component (typically, $\Gamma \sim$ 1.4 -- 2.1) accompanied by weaker
multi-colour black-body emission associated with the accretion
disc, with $kT_{\rm in} < 0.5$ keV \citep{miller2006,reis2010}.  While
the source of black-body photons is strongly suspected to be the
accretion disc, there is some controversy
as to the origin and physical location of the power-law component,
which could be produced by Compton scattering in a hot corona
\citep{malzac2009}, in a hot inner flow \citep{zdziarski1998}
resulting from the evaporation of the inner radii of the accretion
disc (thus truncating the disc, see e.g. \citealt{narayan1994}), or at the base of the radio-emitting jet that is observed during this state
\citep{markoff2005}. Notwithstanding the actual physical source of
hard photons (i.e. $E\gtrsim 2$ keV), there is general agreement that
the hard photon-emitting region must be relatively central, 
within tens of gravitational radii from the central object in the
bright hard state.

Previously, most X-ray variability studies of the hard state were conducted using instruments with hard X-ray sensitivity, e.g. the Proportional Counter Array (PCA) on the {\it Rossi X-ray Timing Explorer} ({\it RXTE}), which limited the study of the variability to the hard, power-law component.  Due to the anti-correlation of rms variability amplitude
with spectral hardness as a source transitions to the disc-dominated
soft state (e.g. \citealt{belloni2005}), a picture arose where the disc itself was intrinsically stable and variability is generated in the power-law emitting region, usually envisioned as an unstable hot inner flow \citep{churazov2001,done2007}.

However, recent work \citep{wilkinson2009,uttley2011} has used the
soft X-ray response and timing capability of the EPIC-pn instrument on
board {\it XMM-Newton} to carry out the first spectral-timing studies
of the disc in the hard state. This work has shown strong evidence, in the form of variability (`covariance') spectra and time-lags, that the disc is intrinsically variable, and that disc variations in fact precede (and likely drive) the power-law variations, at least on time-scales of seconds and longer.  The role of the disc as the driver of variability can be explained by invoking the presence of
mass-accretion rate fluctuations that originate in the disc at
particular radii and propagate inwards. Such fluctuations vary at the
local viscous time-scale corresponding to the radius where they are
produced \citep{lyubarskii1997,kotov2001,arevalo2006}. Because the $\dot{m}$ variations at inner radii are
modulated by the outer $\dot{m}$ via a multiplicative process in the
context of this model,
variability is seen at all frequencies that correspond to the
viscous time-scales of the radii generating the propagating signals.
Once the perturbations reach the inner regions of the system, these
modulate the power-law emission that accounts for most of the 
hard X-ray flux. A direct implication of this model is that
the variations in power-law X-ray emission must be observed after the
disc black-body variations with a time-delay scaling with
the viscous travel-time between the radii where the fluctuations
originate and the power-law emitting region.

So far, evidence for disc-driven variability has been seen in {\it
XMM-Newton} EPIC-pn observations of three hard state sources:
GX~339-4, Cyg~X-1 and SWIFT~J1753.5--0127 \citep{uttley2011}.  These data were obtained
during relatively long-lived hard states at moderate luminosities,
around 1~per~cent of the Eddington limit (with some uncertainty given
the uncertain distances to GX~339-4 and SWIFT~J1753.5--0127 as well as
some uncertainty on their masses).  More luminous hard states, corresponding to outburst rises or `failed' state transitions, have previously not been studied with detailed spectral-timing into the soft X-ray band, but could contain important information about changes in the disc variability as the accretion rate increases and the disc emission strengthens. 
In this paper, we carry out the first detailed spectral-timing study of a luminous hard state of SWIFT~J1753.5--0127, a transient black hole candidate which was
discovered on 2005 May 30 \citep{palmer2005}. Its $\sim
3.2$ hr period derived from optical lightcurves
\citep{zurita2008,durant2009} makes it the BHXRB with the second-shortest period
known to date \citep{kuulkers2012}. Since the start of the outburst, it has
undergone a transition to and from the hard-intermediate state. It has never
completed the outburst cycle toward softer states nor has it gone into
quiescence. The long-term {\it Swift/BAT} lightcurve of
SWIFT~J1753.5--0127 is shown in Figure~\ref{fig:outb} and shows the
epochs of the two {\it XMM-Newton} EPIC-pn observations.  The first, obtained early in 2006, corresponds to a relatively faint hard state and shows clear evidence for disc-driven variability \citep{wilkinson2009,uttley2011}.  The second observation was triggered by us in response to a brightening of the hard state in September 2009, and enables a comparison of a more luminous hard state with the lower-luminosity states studied to date. 

In Section~\ref{sec:datared} we describe our data reduction, and in Section~\ref{sec:results} we carry out a detailed spectral-timing study of the 2009 data and compare our results with the 2006 observation.  In Section~\ref{sec:discussion} we summarise our key results and interpret them in terms of the disc fluctuation model which can explain the data obtained at lower luminosities.

%%%%%%%%%%%%%%%%%%%%%%%%%%%%%%%%%%
\section{Data reduction}
%%%%%%%%%%%%%%%%%%%%%%%%%%%%%%%%%%
\label{sec:datared}

\begin{figure}
\centering
\includegraphics[width=.47\textwidth]{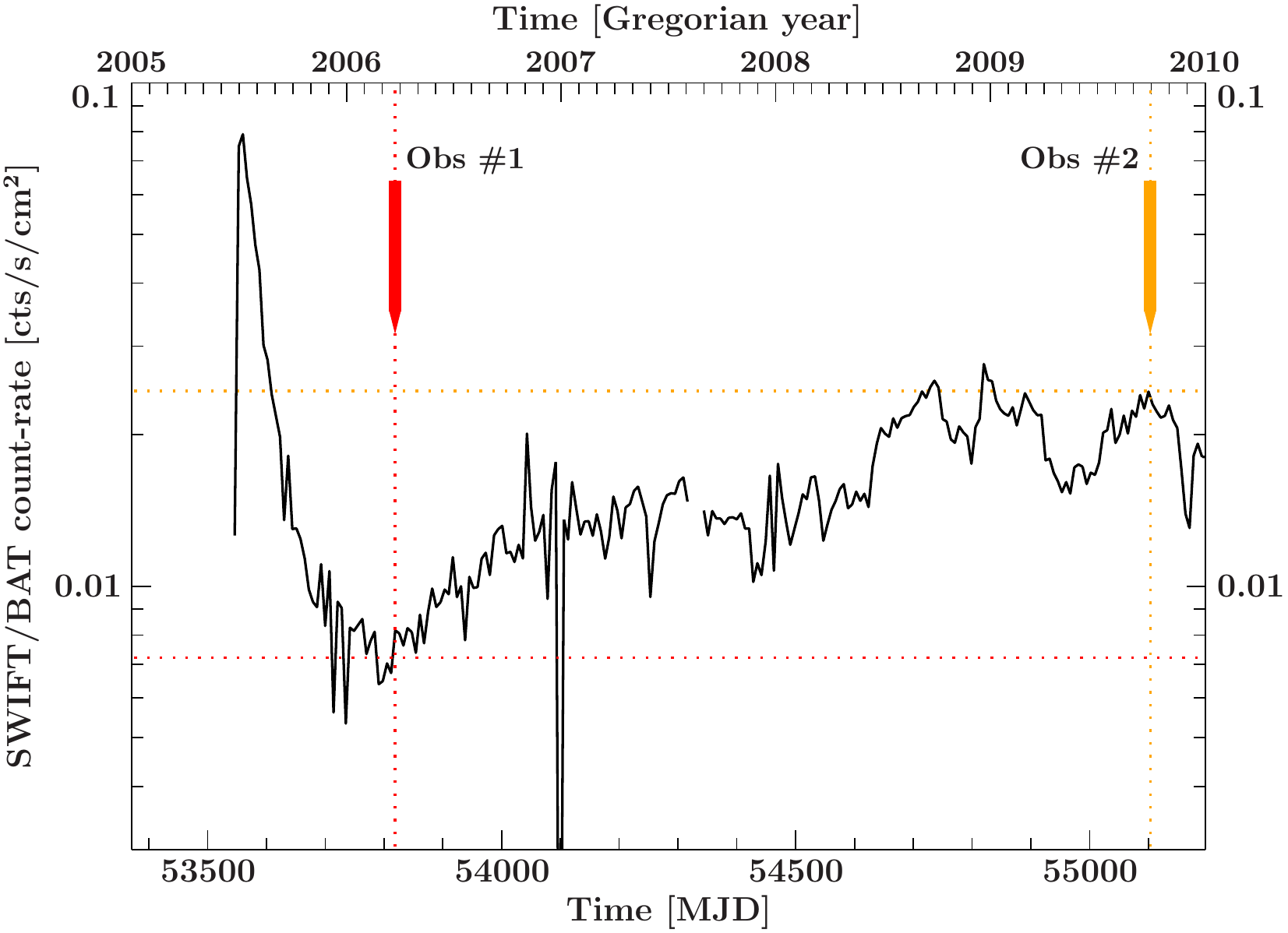}
\caption{{\it Swift/BAT} lightcurve of SWIFT~J1753.5--0127, taking
weekly flux averages. The 2006
(\#1) and 2009 (\#2) {\it XMM-Newton} observations of the source are marked in red
and orange, respectively.}
\label{fig:outb}
\end{figure}

SWIFT~J1753.5--0127 was observed with \textit{XMM-Newton} on 
2009 September 29 (ObsID 0605610301) and is shown as Obs \#2 on
Fig.~\ref{fig:outb}, together with the 2006 observation (Obs \#1,
ObsID 0311590901) that we use later for comparison. We have extracted the data from the EPIC-pn
camera in Timing Mode \citep{kuster2002} using the SAS 10.0.0 tool
\texttt{epproc} and the most recent Current Calibration
Files (CCFs). We have used \texttt{evselect} for filtering and extracted only single and
double events (\texttt{PATTERN <= 4}) during intervals of low, quiet
background. The column filters \texttt{RAWX in [27:50]} was chosen for
the 2006 observation, and the selection \texttt{RAWX in [27:35] || RAWX in
[38:50]} was adopted for the 2009, thus excising the central two
columns to mitigate pile-up effects. Although an excision of further
columns results in a further decrease of pile-up, this effect does not
change any of the results shown in this paper. Due to telemetry
drop-outs, the total useful exposure are $\sim 35.5$~ks (2006) and
$\sim 13.3$~ks (2009).

The tool \texttt{evselect} was used to extract the spectrum and 
the response files were extracted with \texttt{rmfgen} and \texttt{arfgen}.
Power spectra, cross-coherences, phase- and time-lags were extracted using data within the GTI
intervals following \citet{nowak1999} and \citet{vaughan1997}. We use a time bin of 0.005965~s and
segment size of 16384 and 4096 bins/segment (for the 2006 and 2009
observations, respectively).

We omit any segments with gaps between
successive events longer than 0.1~s. These gaps are likely to be due to telemetry drop-outs that are not always included in the GTI files.
Covariance \citep{wilkinson2009} and rms spectra \citep{revnivtsev1999} as well as the energy-dependent
phase- and time-lags were also
extracted for different frequency ranges as discussed in
Section~\ref{subsec:cosp}. 

Given the Fourier transform of the signal in the $i$-th energy channel
(or energy band) $S_i(\nu_l)$, the
cross-spectrum between two bands 1 and 2 is defined as:

\begin{equation}
\langle C_{1,2}(\nu_l)\rangle=\langle S_1^*(\nu_l) S_2(\nu_l) \rangle
\label{eq:cs}
\end{equation}

\noindent where the averages are performed over a number of independent
segments and frequency bins $m$. Energy-dependent products such as phase- and time-lags as well as
covariance spectra can be obtained from the above quantity by
defining a reference band for each channel/band 
$ S_{\bar{i}} (\nu_l) = $
$\sum_{i' \neq i} S_{i'}(\nu_l)$. In reality, the reference band can be fine-tuned to
incorporate a given set of channels (e.g. a soft or a hard band), as we will show in the Results
section.
With this definition, the cross-spectrum for the $i$-th channel becomes:

\begin{equation}
\langle C_i(\nu_l) \rangle= \langle S_{\bar{i}}^*(\nu_l) S_i(\nu_l) \rangle
=\left \langle  \sum_{i' \neq i} S_{i'}^*(\nu_l) S_i(\nu_l) \right \rangle
\label{eq:cs2}
\end{equation}

The energy-dependent phase-lag is then given by $\phi_i(\nu_l) = \arg\big[
\langle C_i(\nu_l)\rangle \big]$ and its time-lag is $\tau_i(\nu_l) =
\phi_i(\nu_l) / 2 \pi \nu_l$.

The covariance spectrum in the frequency range $\nu_{l_1}$ to
$\nu_{l_2}$ can be defined as \citep{wilkinson2011}:

\begin{equation}
\textrm{Cov}_i(\nu_l) = 
\frac{\sqrt{\sum_ {l={l_1}}^{{l_2}} \left ( \langle |
C_i(\nu_l)|^2\rangle 
- B_i(\nu_l)\right ) \Delta \nu_l}}
{\sqrt{\sum_{l={l_1}}^{{l_2}} \langle |
S_{\bar{i}} (\nu_l )|^2 \rangle \Delta \nu_l }}
\label{eq:cov}
\end{equation}

\noindent where $\Delta \nu_l$ is the width of the $l$-th frequency
bin and $B_i(\nu_l)$ denotes the bias caused by the
contribution of Poisson power $N_i$, which is subtracted from the
noiseless modulus-squared value of the cross-spectrum (see e.g.
\citealt{vaughan1997}):

\begin{align}
\begin{split}
B_i(\nu_l) = &  (\langle |S_{\bar{i}}(\nu_l)|^2 \rangle \langle
|S_{i}(\nu_l)|^2
\rangle  + \langle | S_{\bar{i}}(\nu_l)|^2 \rangle  \langle N_{i}(\nu_l) \rangle\\ 
 + &\langle N_{\bar{i}}(\nu_l) \rangle \langle N_{i}(\nu_l)
 \rangle ) / m_l
\end{split}
\label{eq:bias}
\end{align}

The spectral analysis was performed in ISIS 1.6.2-1 \citep{houck2000},
using a spectrum where a minimum of 3 channels per bin are grouped together, after adding 1\% systematic error to account for instrumental
uncertainties in the calibration of the instrument \citep{wilkinson2009}.

%%%%%%%%%%%%%%%%%%%%%%%%%%%%%%%%%%
\section{Spectral-timing analysis and results}
%%%%%%%%%%%%%%%%%%%%%%%%%%%%%%%%%%
\label{sec:results}

\subsection{Spectral analysis}
\label{subsec:specfit}
\begin{figure}
\centering
\includegraphics[width=.45\textwidth]{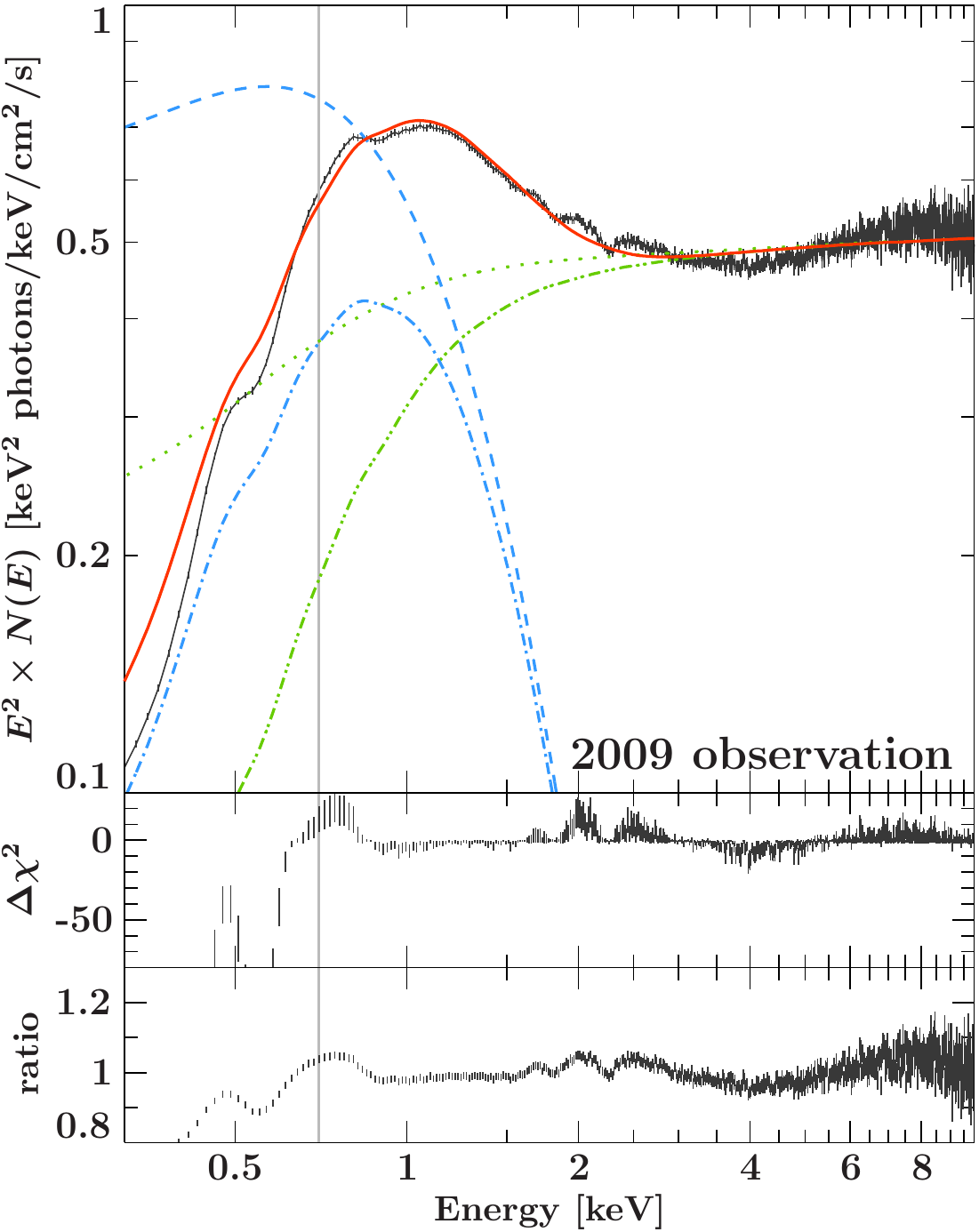}
\caption{2009 spectrum (black) and best-fitting model (red solid line)
using the model
\texttt{tbnew(1)}\texttt{*(diskbb(1)}\texttt{+nthcomp(1))}. The $\Delta \chi^2$ and ratio
residuals are shown in the lower panels. The disc black-body component
is shown in blue (absorbed is shown dash-dotted, unabsorbed is shown dashed),
whereas the Comptonisation component is depicted in green (absorbed is
shown dash-dot-dotted, unabsorbed is shown dashed).
The spectrum below 0.7 keV (left side of the
vertical line) is shown but not used for the fit.}
\label{fig:specfit}
\end{figure}

We first apply a simple phenomenological model to the energy spectrum in order to
understand qualitatively the contributions of the key continuum components, which will inform our choice of energy-bands for spectral-timing analysis.  A cursory examination of the unfolded spectrum reveals a clear soft X-ray excess above the usual power-law emission, consistent with a stronger disc black-body component than is seen at lower luminosities in the hard state.  Therefore we consider a model consisting of a multi-colour disc black-body 
emission component plus a thermal-Comptonisation 
component (\texttt{nthcomp}, hereinafter the `power-law' component
given the similarity in shape at the hard energies), both absorbed by neutral Galactic absorption
with the model \texttt{tbnew}
\footnote{http://pulsar.sternwarte.uni-erlangen.de/wilms/research/tbabs/}
\citep{wilms2011}, assuming the abundances in \citet{wilms2000} and
the cross-sections in \citet{verner1996}.
In ISIS, the model corresponds to \texttt{tbnew(1)}\texttt{*(diskbb(1)+nthcomp(1))}. In addition, we fix the neutral hydrogen column density $N_H$ value to $0.194\times10^{22}$
cm$^{-2}$ \citep{wilkinson2009} and the thermal-Comptonisation
electron temperature $kT_e$ to 53 keV \citep{reynolds2010}. The seed
photon temperature for Comptonisation is tied to the disc black-body
inner disc temperature. The model is fitted only to the 0.7--10~keV
energy range, following the recommendations of the EPIC Consortium
not to consider the data below 0.6~keV \citep{guainazzi2011}, however
we plot the entire useful energy range covered by the data for
completeness and to show clearly the predicted disc contribution.

The best-fitting parameters for the model are shown
in the first column of Table~\ref{tbl:specfit} and the spectrum together with the
best-fitting model is shown in Figure~\ref{fig:specfit}. Although the fitting
procedure yields a high $\chi^2/$d.o.f. $\sim 2.6$ (possibly because
of underestimated instrumental systematic errors), our aim is not to
provide an accurate spectral fit but rather to understand what
emission components are the fundamental spectral sources of photons.
Therefore, we do not include the non-primary spectral components in the fit, such as the disc reflection component.

\begin{figure}
\centering
\includegraphics[width=.45\textwidth]{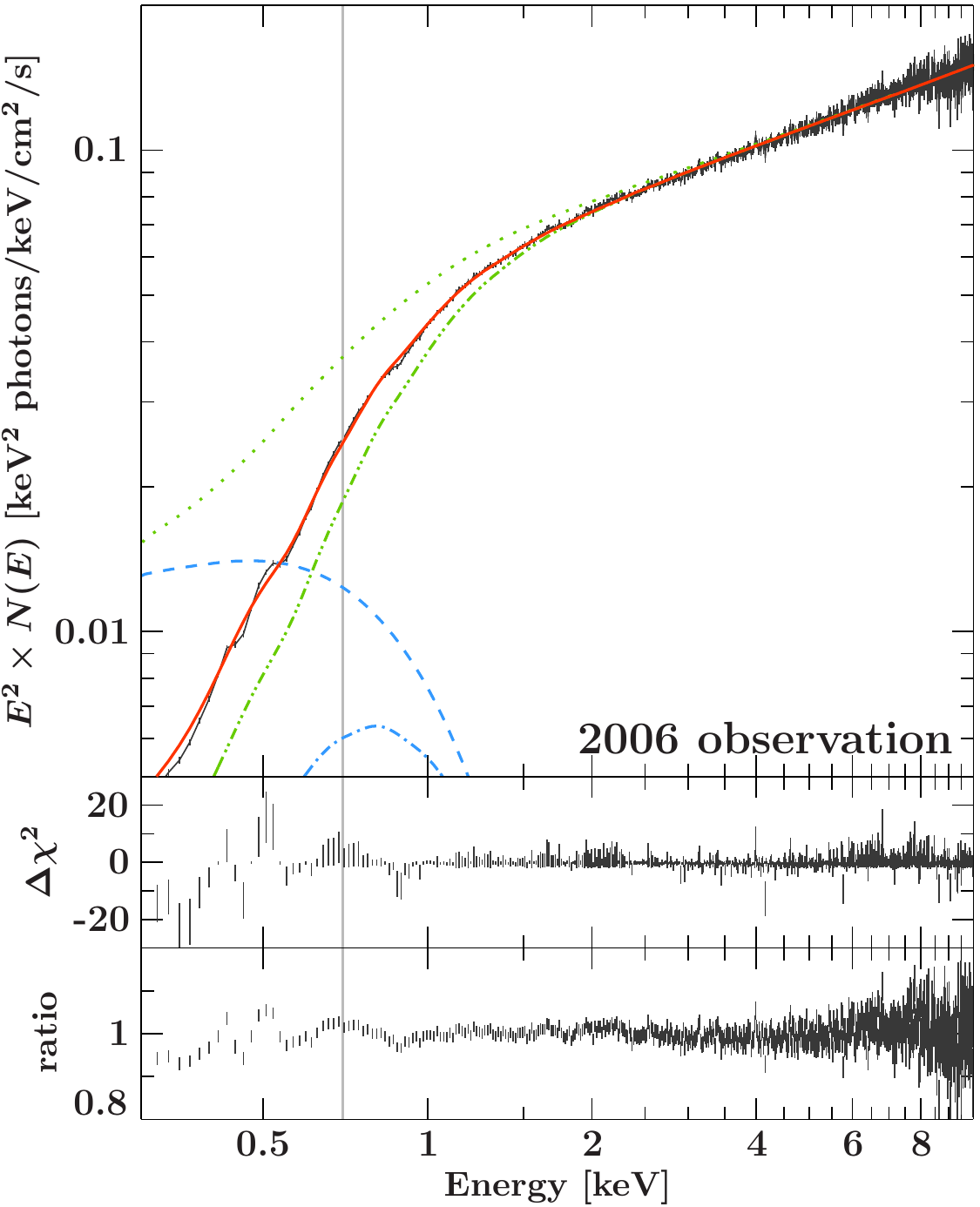}
\caption{Same as Fig.~\ref{fig:specfit}, for the 2006 observation}
\label{fig:specfit6}
\end{figure}

We use the same model to re-fit the EPIC-pn spectrum of the 2006 hard-state observation presented in \citet{miller2006} and
\citet{wilkinson2009}, which we will later use for a comparison with
the 2009 data.  The fitted parameters are shown in
Table~\ref{tbl:specfit} and the model fit is plotted in Fig.~\ref{fig:specfit6}. In
this case we find a harder power-law ($\Gamma = 1.620 \pm 0.005$) and
a  much weaker disc black-body component with a lower
temperature ($kT_{\rm in} = 0.212^{+0.020}_{-0.018}$) and
normalisation only $\sim3$~percent  of the value from 2009.
\begin{table}
\caption{Best-fitting values for the model
\texttt{tbnew(1)*(diskbb(1)+nthcomp(1))} fitted over the range 0.7 --
10.0 keV. The disc black-body
normalisation $N_{\rm dbb}$ is in units of $[(r_{\rm in}/{\rm km}) / (D / 10 {\rm
kpc})]^2 \cos \theta$ (where $r_{\rm in}$ is the `apparent' disc inner
radius that is also dependent on the inclination $\theta$ of the disc,
see \citet{kubota1998}), whereas the
Comptonisation normalisation is in units of ${\rm photons~cm}^{-2} {~\rm
s}^{-1} {~\rm keV}^{-1}$ at 1 keV.}
\centering
\begin{tabular}{lcc}
Parameter & Value (2009) & Value (2006)\\
\hline
% ad.sl
$N_H$ [$10^{22}$ cm$^{-2}$] & 0.194 (fixed) & 0.194 (fixed) \\
$kT_{\rm in}$ [keV]  & $0.248\pm 0.001$ & $ 0.212^{+0.020}_{-0.018}$\\
$kT_{\rm e}$ [keV]  & 53 (fixed) & 53 (fixed) \\
$N_{\rm dbb}$ & $31490^{+810}_{-940}$ & $ 1050^{+270} _{-50} $\\
$\Gamma$ & $1.961^{+0.004}_{-0.005}$ & $1.620 \pm 0.005$ \\
$N_{\rm nth}$ & $0.429\pm 0.003$ & $0.052 \pm 0.003$\\
\hline
$\chi^2$/d.o.f. & $1638/617$ & $600/623$ \\
\hline
\end{tabular}
\label{tbl:specfit}
\end{table}

\subsection{2009 data: frequency-dependent spectral-timing properties}
\label{subsec:timing}
\begin{figure}
\centering
\includegraphics[width=.45\textwidth]{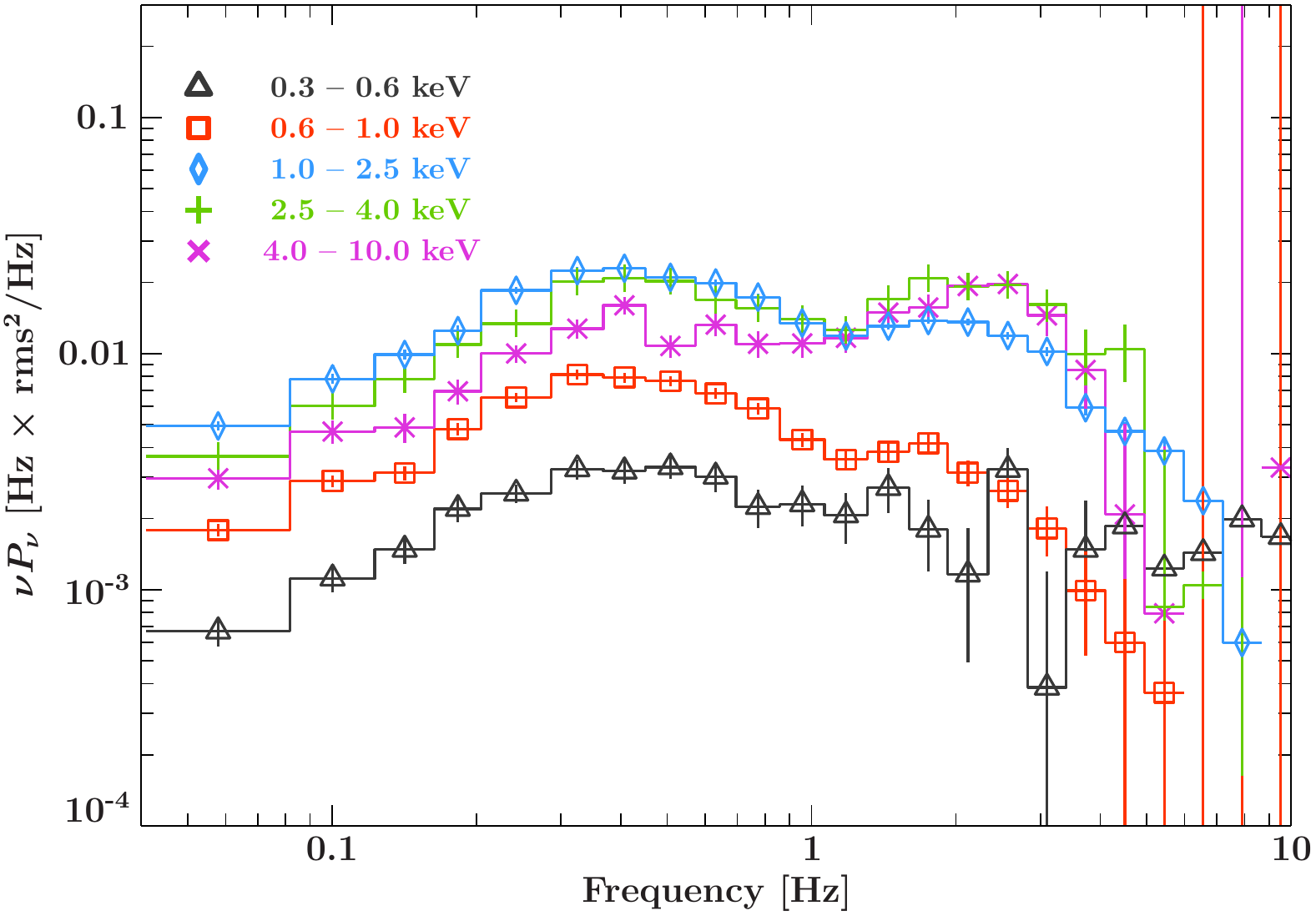}
\caption{Power spectral densities (PSDs) for the 2009 observation in the energy bands 0.3 --
0.6 keV (black triangles), 0.6 -- 1.0 keV (red squares), 1.0 -- 2.5
keV (blue diamonds), 2.5 -- 4.0 keV
(green plus signs) and 4.0 -- 10.0 keV (magenta times signs) showing two broad peaks whose
relative amplitude varies as a function of the energy band chosen.}
\label{fig:psd}
\end{figure}

\begin{figure*}
\centering
\includegraphics[width=\textwidth]{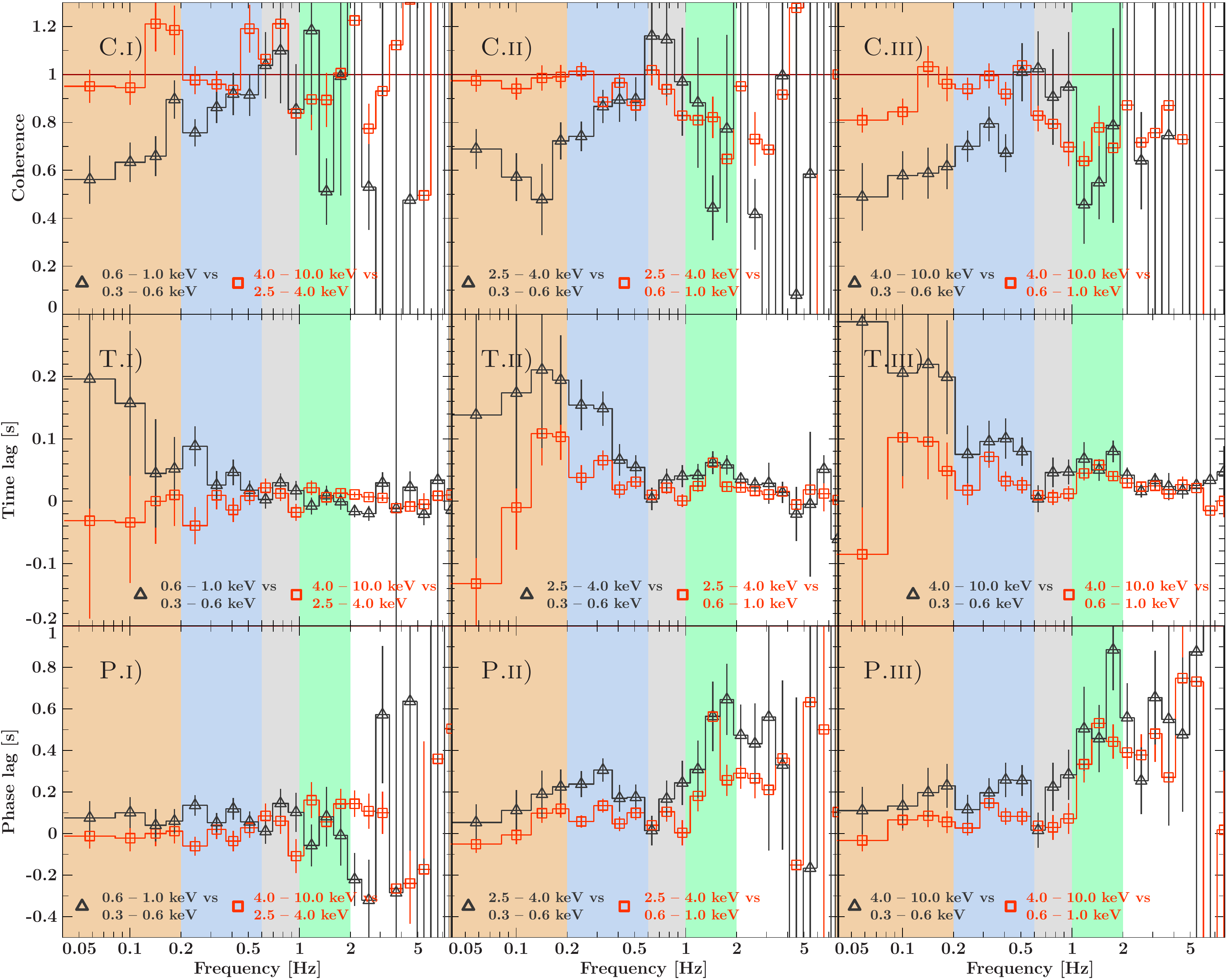}
\caption{Coherence, time- and phase-lags for the 2009 observation,
using different combinations of the bands 0.3 -- 0.6 keV, 0.6 -- 1.0
keV and 4.0 -- 10.0 keV. 
A positive lag indicates that the emission in the harder band lags the
emission in the softer band.
The shaded regions indicate the frequency intervals used to average
the energy-dependent products studied in Section~\ref{subsec:cosp}.}
\label{fig:freqprods2}
\end{figure*}

The presence of both a disc black-body component and a Comptonisation
component that dominate different energy bands covered by EPIC-pn, as seen
in Figure~\ref{fig:specfit}, suggests that the variability properties
of the source may differ where the photon flux is dominated by either
the disc black-body or the power-law components.  Hence it 
is convenient to extract light curves for the energy ranges 0.3
-- 0.6 keV, 0.6 -- 1.0 keV, 2.5 -- 4.0 keV and 4.0 -- 10.0 keV,
hereinafter ultrasoft, soft, intermediate and hard bands, respectively. For completeness we also
extracted data from the the 1.0 -- 2.5 keV band, although since
this band shows no unique properties compared to the other bands, we
do not show it in our later spectral-timing results, but we do include it in showing the energy-dependent power-spectral densities (PSDs), which are shown in Fig.~\ref{fig:psd}.  As shown in Fig.~\ref{fig:specfit}, the
ultrasoft and soft bands are dominated by the disc black-body
component, whereas the intermediate and hard bands cover the
power-law component. The range 1.0 -- 2.5 keV represents
the variability power for the photons emitted by a combination of the
black-body and power-law components.

In addition, the ultrasoft, soft, intermediate and hard bands are also
used to extract the coherence, phase- and time-lags, using the
combinations 
ultrasoft-soft, ultrasoft-intermediate, ultrasoft-hard,
soft-intermediate, soft-hard and
intermediate-hard, shown in Fig.~\ref{fig:freqprods2}. The properties of
coherence and lags with respect to the 1.0 -- 2.5 keV band show an
intermediate behaviour compared to other bands and are not shown here.

\subsubsection{Amplitude of variability}

The power spectral densities in Fig.~\ref{fig:psd} 
show evidence for two broad noise components, peaking at $\sim 0.4$ and $\sim 2$ Hz
(hereinafter, the `low-frequency' and `high-frequency' peaks respectively).  The relative amplitude
between the two peaks changes, depending on the energy band considered.
At the soft and
ultrasoft energies where the disc black-body emission dominates, the
low-frequency peak shows a larger peak-amplitude of variability than the high-frequency peak.
Conversely, for the power-law-dominated intermediate
and hard energies the high-frequency peak-amplitude becomes stronger with respect to the low-frequency peak and is
consistent between the two bands, while the 1.0 -- 2.5 keV band shows
an intermediate behaviour between the softer and the higher-energy bands.  Besides the energy-dependence of the PSD shape, there is also a clear change in normalisation, with the ultrasoft band showing the weakest variability at all frequencies, followed by the soft band and then the higher-energy bands.  This behaviour contrasts with that seen in the 2006 data by \citet{wilkinson2009}, where the softer 0.5-1~keV band showed {\it larger-amplitude} variability than the 2-10~keV band at low frequencies.

\subsubsection{Coherence and frequency-dependent spectra}

The coherence is an indicator of the degree of linear correlation between variations in the
emission in two different bands \citep{vaughan1997} at a particular
frequency. Full coherence (when the coherence equals
unity at the given frequency) is reached when the variations in emission in one band can be reconstructed
linearly from the variations in emission in the other band.  The upper panels in Fig.~\ref{fig:freqprods2} show the coherence computed for all
possible combinations of the above bands. For all plots in the figure (including
time- and phase-lags), the left
panels show the frequency-dependent products computed between the pair
of bands that cover the disc black-body region or the pair of bands which cover the
power-law region, i.e. the disc vs. disc behaviour and the power-law vs. power-law behaviour.  The remaining panels show the
same products made from intercombinations of the disc and power-law bands, combining the intermediate and hard bands (middle
and right panels, respectively) together with the ultrasoft and
soft bands that are dominated by the disc black-body emission.

In a simple case where all of the emission in one spectral
component varies linearly together at all energies, the coherence
between the observed emission in two bands where that component dominates
should approach unity. This is similar to the case observed with the
coherence between the intermediate and hard bands that sample the
power-law (\cri), as well as between the intermediate (\crii) and hard bands
(\criii) with respect to the soft band, up to $\sim 0.6$ Hz. This
frequency roughly corresponds to the point where the low and high-frequency peaks in the
PSD intersect (Fig.~\ref{fig:psd}). The coherence deviates slightly
from unity for frequencies below $\sim 0.1$ Hz between the hard and
soft bands.

A very different behaviour is observed between the ultrasoft and the
soft, intermediate and hard bands. In this case, unity coherence
is reached at $\sim 0.6$~Hz, and decreases towards {\it both} lower and higher
frequencies, reaching values of $\sim 0.5$ in some cases.  Interestingly, the ultrasoft variations at low frequencies are just as incoherent with the soft variations as they are with the hard or intermediate variations.  Thus, at low frequencies, while the variations in the two power-law bands are coherent with each other, a significant fraction of the variations in the two disc bands are incoherent with each other.

\subsubsection{Time and phase lags}
\label{subsec:timephaselags}

Previous observations of time-lags in the hard state show that they roughly follow a frequency-dependence $\tau(\nu) \propto
\nu^{-0.7}$ \citep{miyamoto1989,miyamoto1992,crary1998,nowak1999},
albeit with detectable `steps' in the lag vs. frequency dependence
where the data are good enough to discern them \citep{nowak1999}.  For
the observation that we are studying, the frequency dependence does
clearly
vary as a function of energy bands chosen. For lags between the
two disc-dominated bands (ultrasoft and soft) or between the intermediate and
hard power-law bands, a relatively small amplitude is seen with a weak and smooth evolution of phase lag with Fourier-frequency.  However, the lags between power-law and disc bands show a much more pronounced change with a significant `step' (most clearly visible in the phase lags) above 1~Hz.  Thus the strongest evolution of lags with frequency occurs when comparing the lags between the power-law and disc bands, with the power-law lagging the disc variations, as seen in other observations of hard state BHXRBs \citep{uttley2011}.

\subsection{2009 data: energy-dependent lags and covariance spectra}
\label{subsec:cosp}

\begin{figure*}
\centering
\includegraphics[width=0.98\textwidth]{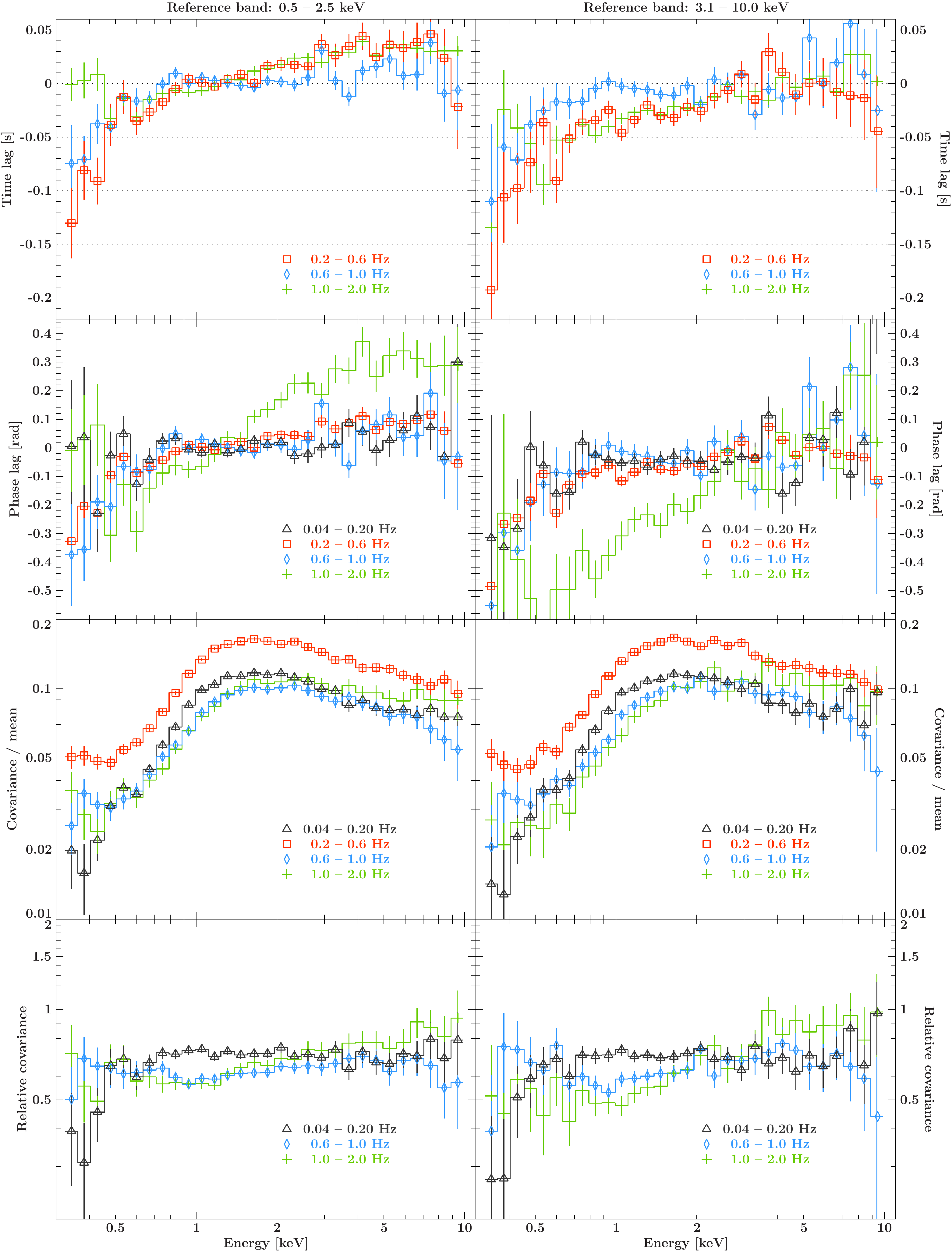}
\caption{Time- and phase-lag spectra, covariance spectrum divided by
average spectrum and covariance spectrum divided by covariance
spectrum computed in the range 0.2 -- 0.6 Hz for 2009 observation. The frequency ranges
used are 0.04 -- 0.20 Hz (black triangles), 0.2 -- 0.6 Hz (red squares), 0.6 -- 1.0
(blue diamonds) and 1.0 -- 2.0 Hz (green plus signs). The left and
right plots show the products above using the 0.5 -- 2.5 keV and 3.1
-- 10.0 keV reference bands, respectively.}
\label{fig:enprods}
\end{figure*}

Given the complex behaviour in the coherence and lags shown in Fig.~\ref{fig:freqprods2},
we next pursue an energy-resolved analysis that can more clearly relate this behaviour to the different spectral components that contribute to the
variability of the source.

We extract cross-spectra in the frequency ranges 0.04 -- 0.20
Hz, 0.2 -- 0.6 Hz, 0.6 -- 1.0 Hz and 1.0 -- 2.0 Hz (that correspond to
each of the shaded regions in Fig.~\ref{fig:freqprods2}) that
correspond to frequency ranges with comparable behaviour in terms of lags and coherence.
The cross-spectra are computed between a
broad, high signal-to-noise reference energy band, and each of 30 rebinned
energy channels. Following \citet{wilkinson2009} and \citet{uttley2011}, the reference band light curve is determined separately for
each channel since it must have the signal in the channel of interest subtracted
in order to avoid correlated Poisson noise effects. In these
previous works, the data were largely coherent at all
frequencies, so the results were not sensitive to the choice of
reference band. In the observation presented here, our spectral-timing products could vary with the choice
of reference band, because the lags and covariance spectra are
effectively weighted by the coherence of each channel with the
reference band.  Therefore, to account for any effects due to the
choice of reference band, we make cross-spectra for two reference
bands: 0.5 -- 2.5 keV and 3.1 -- 10.0 keV.  Following
\citet{uttley2011}, we make energy-dependent lag and covariance spectra for each of these two bands, plotted as the left and right columns in Fig.~\ref{fig:enprods}.

\subsubsection{Lag-energy spectra}
The time and phase-lag versus energy are shown in the top two sets of
panels in Fig.~\ref{fig:enprods}.  Because the lag difference between
two separate energy channels is relative to the choice of reference
band, the offset on the $y$-axis is not meaningful in this
case.  The main consideration is the relative lag between channels, which is plotted here so that more positive values of lag are lagging smaller/more-negative values.  We do not plot the time lags for the lowest-frequency range, since their large values would make it difficult to read the data for the other frequency ranges (these data can be seen in the phase lag plots, however).

The dependence of the lags on energy can be approximated as a
`broken' log-linear law, with a steeper slope below $\sim 1.0$ keV and
a flatter dependence above that energy. This turning point is close to
the energy where the photon flux from the Comptonisation component
overcomes the disc black-body photon flux (Fig.~\ref{fig:specfit}). In the frequency range 0.2 -- 0.6 Hz, the lag
between the power-law and disc components reaches values up to $\sim
0.2$ s, reaching values close to zero as the frequency increases. This
behaviour was also shown in \citet{uttley2011} and is consistent with
the disc variations leading (and probably driving) the power-law
variations on time-scales of seconds.  At frequencies of 1 -- 2~Hz, the low-energy down-turn in lags is replaced by a small up-turn, also similar to what was observed for other observations of BHXRBs at similar frequencies \citep{uttley2011}.  However, it is interesting to note that in this frequency range, the log-linear shape appears also to break downwards at around 2 keV (this behaviour is seen most clearly in the phase lag plot), suggesting that there is a composite shape, perhaps consisting of leading higher-energy disc emission together with lagging lower-energy disc emission.  It should be noted that the general behaviour in the lags is replicated using both choices of reference band, albeit with small differences between them. 

\subsubsection{Covariance spectra}
In addition to the causal information shown above, we extract
covariance spectra from the cross-spectra
\citep{wilkinson2009,uttley2011} in order to quantify, as a function of energy, the amplitude of
variable emission that is
linearly-correlated with the reference band of choice (see
Section~\ref{sec:datared}).
The covariance spectrum depends on the coherence between the reference
band of choice and each energy channel. In the limit where the
coherence is unity, the covariance spectrum is equal to the rms
spectrum (e.g. \citealt{revnivtsev1999}),
although with considerably smaller error-bars due to the effect of the broad reference band acting as a matched filter on the flux variations in individual channels.

The panels in the third row of Fig.~\ref{fig:enprods}
show the corresponding covariance spectra divided by the mean
spectrum, which enables us to estimate the contribution of each component to the overall variability, analogous to the fractional rms spectrum. 
For energies above $\sim 2$ keV, the fractional covariance
spectra show a roughly power-law slope that becomes harder for increasing
frequencies. At the lowest frequencies, the power-law slope is softer
than the mean, while it reaches the slope of the average spectrum at the
highest frequencies, 1.0 -- 2.0 Hz. This effect is mild if only the
first three frequency ranges are considered, and becomes more obvious
for the 1.0 -- 2.0 Hz range. 

Such an effect can also be seen in the lowest panels of Fig.~\ref{fig:enprods},
by computing the ratio between the covariance in the frequency bands
0.04 -- 0.20 Hz, 0.6 -- 1.0 Hz and 1.0 -- 2.0 Hz and the covariance in
the range 0.2 -- 0.6 Hz. In this case, the variation in power-law slope is
plotted independently of any constant component, which would not contribute to the covariance spectra used to normalise the data.  Thus the effect cannot be due to any simple change in the ratio of a variable to constant spectral component which have different spectral slopes.  Instead, the hardening of the power-law variations at higher frequencies is intrinsic to the variable power-law itself.   Similar behaviour has been seen in frequency-dependent rms spectra of hard state BHXRBs obtained from {\it RXTE} data (e.g. \citealt{revnivtsev1999}).

At the disc-dominated energies (i.e. below $\sim 1$ keV), the fractional covariance 
is up to a factor $\sim 5$ weaker than at the power-law-dominated energies,
suggesting that, although the disc variability is apparently driving the power-law
variability (as seen in the causality argument provided with the lags
above, and in \citealt{uttley2011}), the disc emission is less variable than the power-law emission.  

\begin{figure*}
\centering
\includegraphics[width=\textwidth]{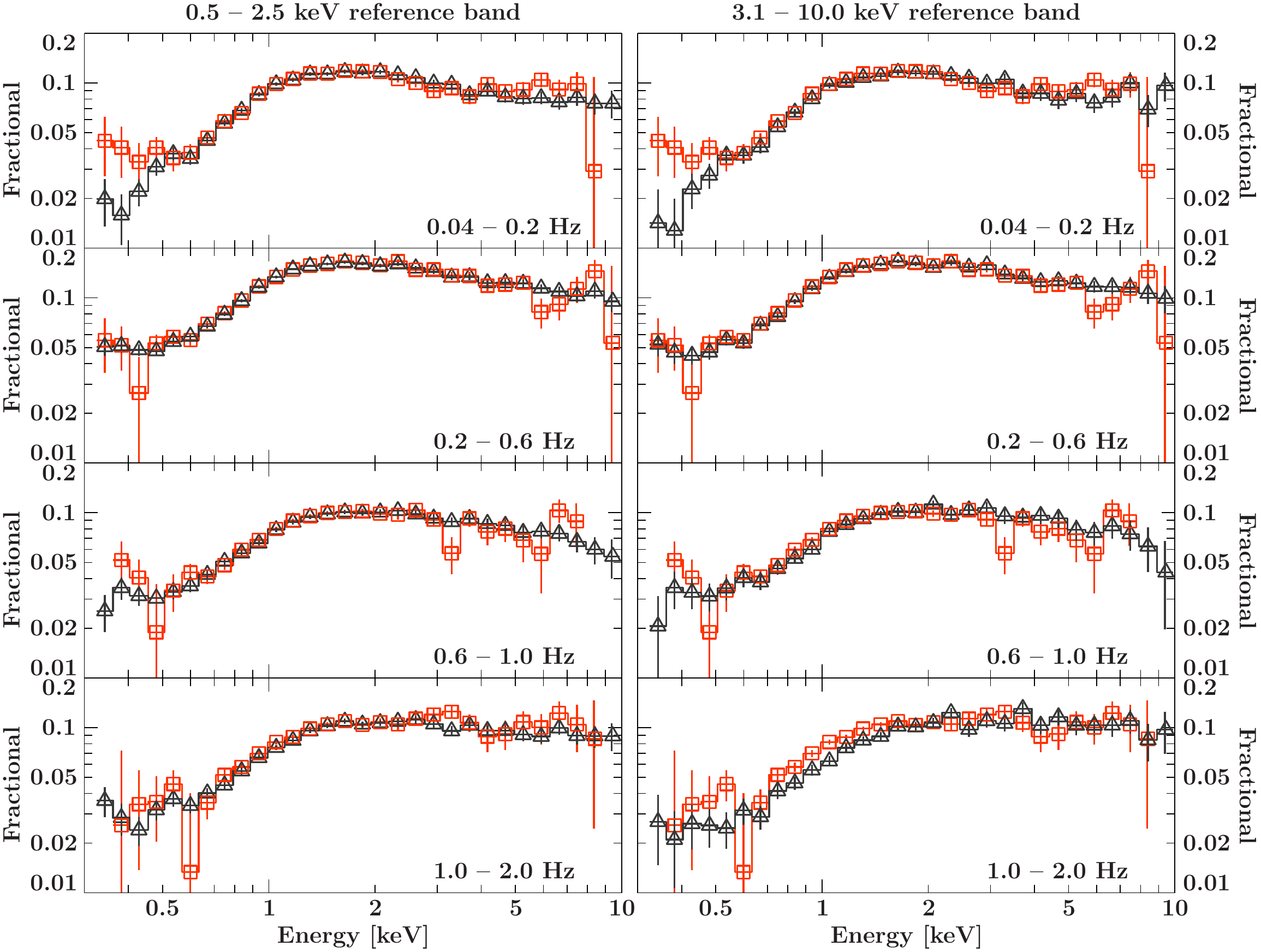}
\caption{Fractional covariance (black triangles) and fractional rms
(red squares) in the
2009 observation for the frequency ranges 0.04 -- 0.20 Hz, 0.2 -- 0.6 Hz, 0.6 --
1.0 Hz and 1.0 -- 2.0 Hz. The left and
right plots show the products above using the 0.5 -- 2.5 keV and 3.1
-- 10.0 keV reference bands, respectively.}
\label{fig:rmsvcov}
\end{figure*}

In Fig.~\ref{fig:freqprods2} we showed that the coherence clearly drops below unity at low frequencies in
any of the combinations of energy bands studied where the ultrasoft
band was considered. As in the previous cases shown above, this drop
can be quantified in an energy-dependent manner by comparing the rms
and covariance spectra.  The fractional covariance spectra shown in Fig.~\ref{fig:enprods} appear to flatten off at the lowest energies in the intermediate frequency bands, while in the lowest-frequency band (where coherence is low) it continues to drop.
We can show that this effect is due to the low coherence of variations of the softest photons by comparing the fractional covariance spectra with the fractional rms spectra, shown in Fig.~\ref{fig:rmsvcov}.  Where the fractional covariance and rms are the same, the coherence is consistent with unity.  However, at low frequencies, the fractional covariance is clearly lower than the fractional rms below $\sim0.5$~keV, indicating sub-unity coherence at these energies, consistent with our findings from the frequency-dependent coherence.  

On the other hand, above 0.5~keV the fractional rms and covariance are
similar for the three lowest frequency ranges, indicating that the
higher-energy part of the disc emission varies coherently with the
power-law emission on time-scales longer than a second.  However, as
also expected from the frequency-dependent coherence measurements, the
coherence of this soft-band emission with the higher energy emission
also drops in the 1--2~Hz frequency bin.  This drop manifests itself as a drop in fractional covariance compared to fractional rms, which is seen at low energies for the 3.1--10~keV reference band, and at high energies for the 0.5--2.0~keV reference band (since this band contains the disc emission).

\subsection{Comparison with the spectral-timing properties in the 2006
hard state}

\begin{figure}
\centering
\includegraphics[width=.45\textwidth]{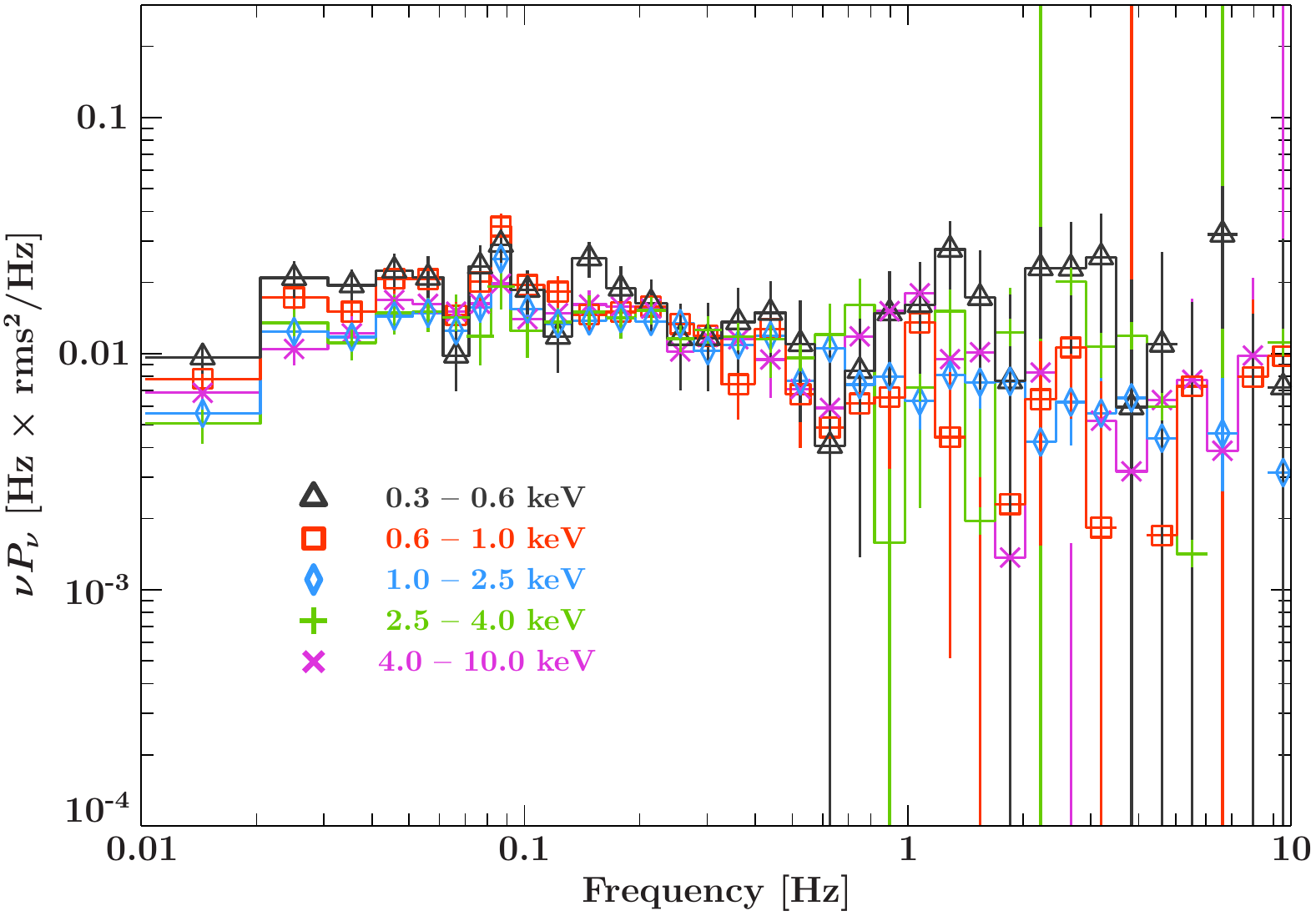}
\caption{Power spectral densities as in Fig.~\ref{fig:psd}, for the
2006 observation}
\label{fig:psd06}
\end{figure}

\begin{figure*}
\centering
\includegraphics[width=\textwidth]{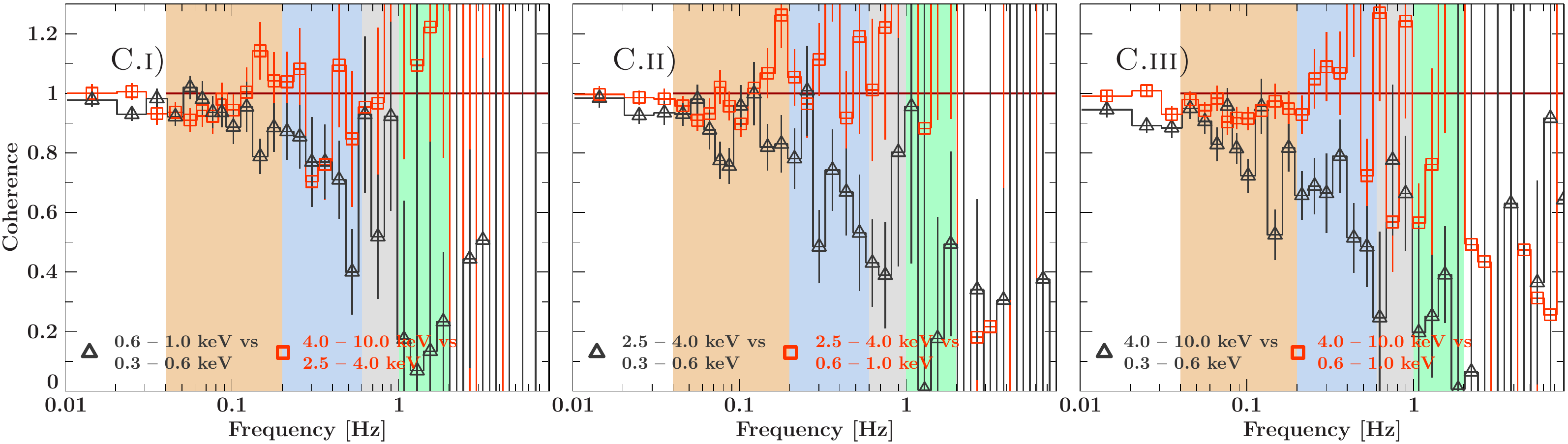}
\caption{Coherence values for the 2006 observation, using the same energy bands
as in Fig.~\ref{fig:freqprods2}}
\label{fig:freqprods206}
\end{figure*}

The 2006 observation was obtained during a 
hard state that was a factor of $\sim 3.4$ fainter than the 2009
observation, as seen in the \textit{Swift/BAT} lightcurve
(red arrow in Fig.~\ref{fig:outb}).  The luminosity difference is even
greater at the lower energies covered by {\it XMM-Newton}, since the
2006 data show a much harder power-law component ($\Gamma = 1.620\pm
0.005$) which dominates the SED, since the disc component is much
weaker (see Figs.~\ref{fig:specfit} and \ref{fig:specfit6} and
Table~\ref{tbl:specfit} for a comparison). Due to the relatively low signal-to-noise in the 2006 data compared to 2009, we limit our comparison of the data to a few key aspects (such as the PSD and coherence) and the two lower frequency ranges for the energy-dependent comparison of lags and covariance.

The power spectral densities for 
the 2006 observation are
shown in Fig.~\ref{fig:psd06}. In this observation, the data can be approximated by a power-law down to $\sim 0.04$ Hz with a similar slope and amplitude for all 
energy ranges considered, as opposed to the much clearer energy dependence of the
power amplitude of the two peaks observed in 2009, whose relative
amplitudes also changed with energy.  The lower-frequency peak in the
PSD seen in 2006 is consistent with the correlation of PSD changes
with spectral evolution seen in moderate to high luminosity hard
states of other BHXRBs, with lower-frequency features seen when the
spectrum is harder (e.g. \citealt{belloni2005}, also seen in
\citealt{boeck2011}).

In the 2009 observation, an important property of the coherence
at frequencies below $\sim 0.6$ Hz is a drop down to coherence values $\sim
0.5$ for each combination of bands that involved the ultrasoft band.  
The corresponding coherence plot for 2006 (Fig.~\ref{fig:freqprods206})
shows a value much closer to unity down to the lowest-frequencies probed.

\begin{figure}
\centering
\includegraphics[width=.46\textwidth]{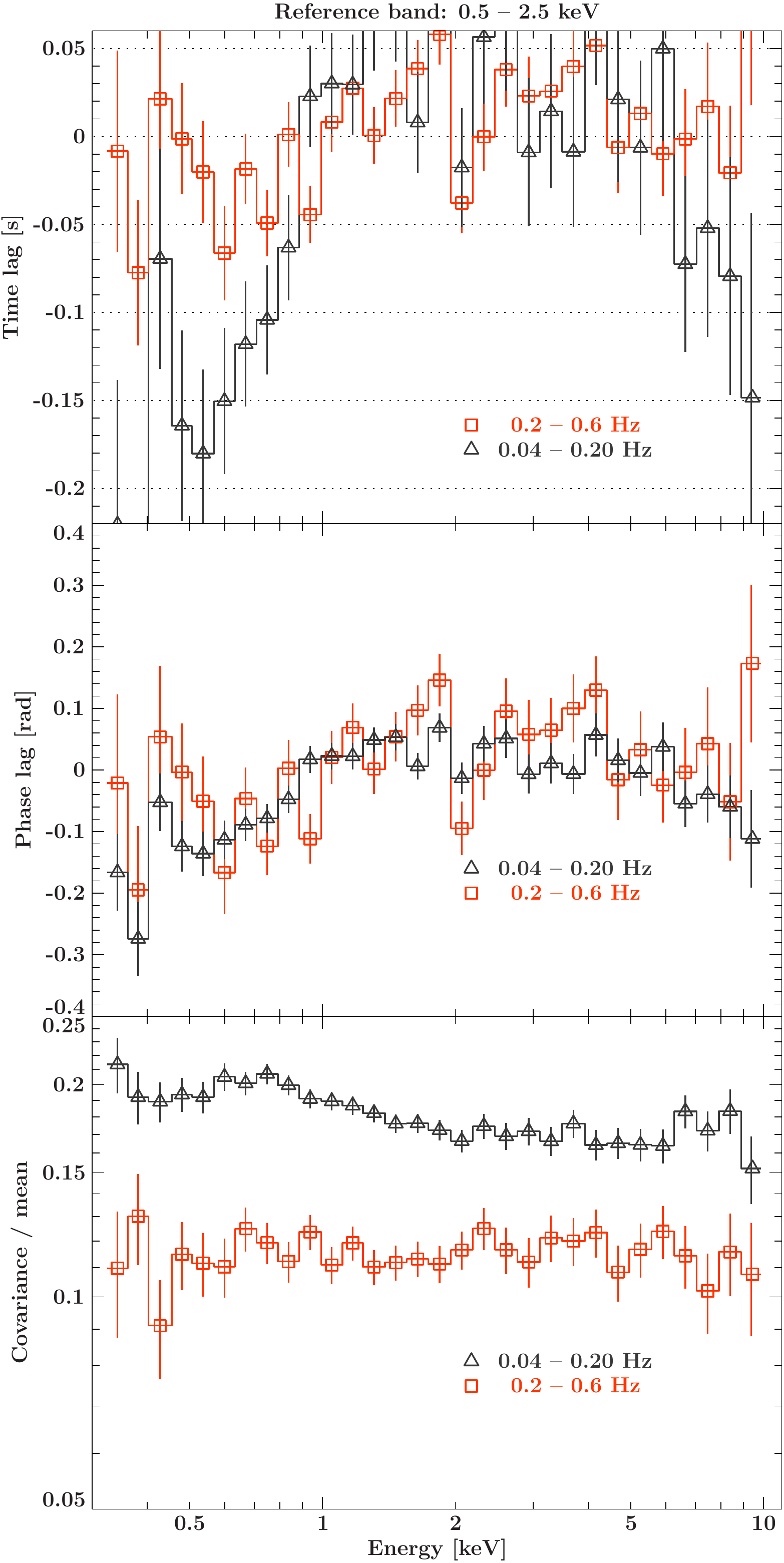}
\caption{Time- and phase-lag spectra and covariance spectrum divided by
average spectrum for 2006 observation, using the 0.5 -- 2.5 keV
reference band. The frequency ranges
used are 0.04 -- 0.20 Hz (black) and 0.2 -- 0.6 Hz (red).}
\label{fig:enprods06}
\end{figure}

Fig.~\ref{fig:enprods06} shows the energy-dependent lags and fractional
covariance spectra for the 2006 data for the first two frequency-ranges (0.04 --
0.20 Hz and 0.2 -- 0.6 Hz), using the reference band 0.5 -- 2.5 keV. The
harder reference band gives consistent but noisier results (as
expected, since the coherence is high at these frequencies). These plots show behaviour consistent with that previously reported for this observation by \citet{wilkinson2009} and \citet{uttley2011}.  The lowest-frequency lags show the characteristic drop below 1~keV that can be associated with the disc driving the variability at harder energies, consistent with the behaviour also seen in 2009 (and implying that there is intrinsic disc variability in both observations).  However, the covariance spectra show a `soft excess' at the lowest frequencies, implying that there is {\it extra} disc variability relative to the power-law variations at these frequencies.  This behaviour contrasts with that seen in 2009 where the disc variations at all frequencies appear to be suppressed when compared with the power-law variations.

\subsection{The emissivity profiles}
\label{subsec:emissivity}

The spectral fits to the 2009 data show much stronger disc emission
than is seen in 2006 (Table~\ref{tbl:specfit}). Although the inferred
disc temperatures are relatively similar, the normalisations (which
are proportional to the square of the inner disc radius) are very different.

Taken at face value, the disc
black-body normalisation in 2009 implies an inner disc radius 5--6
times larger than that in 2006, which is at odds with the standard
interpretation of the other phenomenological changes between the two
data sets, which assumes that the disc becomes less truncated as PSD
characteristic frequencies increase and the disc emission becomes
stronger.  However, to interpret changes in disc black-body
normalisation in terms of changes in inner disc radius, we must assume
in that both data sets the disc emissivity profile is the same.  Since
the X-ray luminosity in 2006 is dominated by the power-law emission,
it is likely that much of the disc emission is produced by X-ray
heating of the disc by the power-law, which could produce a much more
centrally peaked (and apparently `smaller') disc black-body component
(e.g. see \citealt{gierlinski2008}).  In 2009, the disc emission is
comparable to or even exceeds that from the power-law and most of the
emission is likely to be intrinsic to the disc, leading to a more
`standard' emissivity profile. Since the peak energy of the disc black-body
emission is comparable to the low energy end of the spectrum included
in the fit, distinguishing among different emissivity
profiles from the spectral fits would be an arduous task. Systematic
errors deriving from spectral calibration do not, however, impact on
the variability analysis performed here.

Assuming that the disc emissivity profile in 2009 has the standard
$R^{-3}$ form (in terms of bolometric emission) and hence the disc
temperature scales with radius as $R^{-3/4}$, we can consider the
dependence of cumulative disc emission on radius.\footnote{In their
optical/UV-emitting regions, the temperature profiles of irradiated
discs should be flatter than those of disks which are purely viscously heated, if the additional heating
increases the scale-height of the disc (e.g. \citealt{vrtilek1990}) or if the disc
is significantly warped. However, the inner, soft X-ray emitting
regions of discs considered in this paper are likely to be heated
primarily by viscous heating, with, in the case of the 2006 data, an
additional contribution from irradiation which is concentrated in the
centre (i.e. assuming a compact corona). Outside this innermost
region of intense irradiation, the discs are likely to have
temperature profiles closer to those predicted from the viscous
heating models, only flattening on larger scales where soft X-rays are
not emitted.}

\begin{figure}
\centering
\includegraphics[width=.46\textwidth]{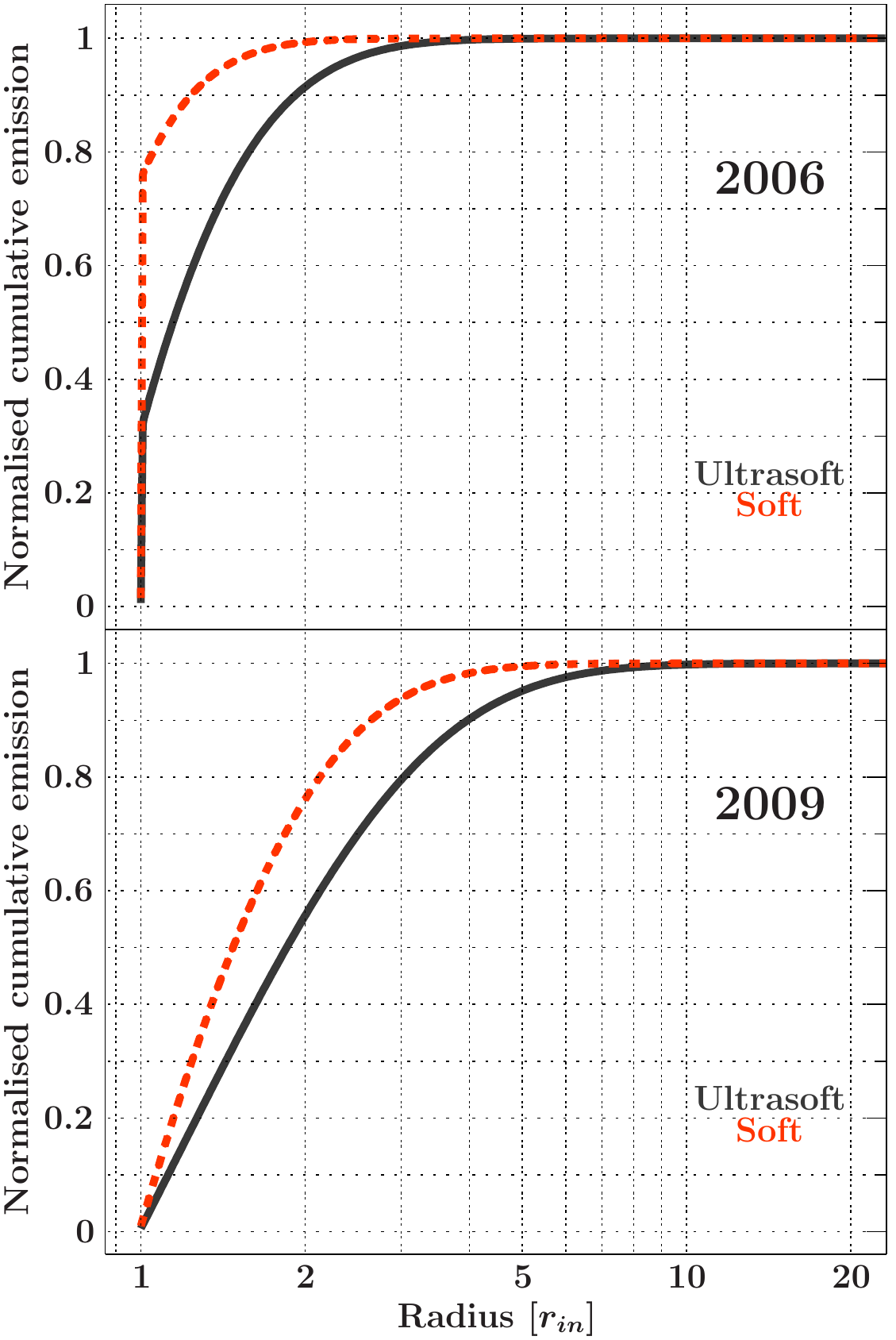}
\caption{Normalised cumulative emission function in the ultrasoft
(solid black) and soft bands (dashed red) as a function of disc
radius. The top panel corresponds to the 2006 data, assuming a composite disc black-body and inner hotter black-body associated with X-ray heating by the power-law emission (see text for details). The bottom panel corresponds to the 2009
data, assuming a dominant intrinsic disc black-body profile.}
\label{fig:emissivity}
\end{figure}

For comparison and allowing for the likely possibility of substantial
disc-heating  by the power-law in 2006, we fit the 2006 spectrum
with a combined single black-body (representing a hot inner ring
heated by the central power-law emission) and a standard disc
black-body component.  We force the disc black-body to have the same
normalisation (i.e. the same assumed inner radius) as the disc in 2009
(note that due to degeneracy in the model the inner radius in 2006 may
be even larger, as expected if the disc is more truncated). For this
composite model the disc black-body temperature drops to 0.11~keV
while the inner single black-body temperature is high at 0.31~keV, but
corresponds to only a small emitting area at the inner edge of the
disc.  The best-fitting model parameters give
$\chi^{2}$/d.o.f.$=565/623$, providing an even better fit to the data
than the single disc black-body model. The resulting cumulative
emission profiles are shown in Fig.~\ref{fig:emissivity} for the soft
and ultrasoft energy bands (after allowing for the effects of the
instrumental response on each band).  

%%%%%%%%%%%%%%%%%%%%%%%%%%%%%%%%%%
\section{Discussion}
%%%%%%%%%%%%%%%%%%%%%%%%%%%%%%%%%%
\label{sec:discussion}

To summarise, our main observational findings are:
\begin{enumerate}
\item  The 2009 {\it XMM-Newton} observation of SWIFT~J1753.5--0127
took place when the source was in a significantly more luminous hard
state ($\sim 6$ times more luminous in the 0.5--10.0~keV band) than in
2006.  Correspondingly, the power-law spectral component is softer
than in 2006, and importantly the soft excess associated with the disc
black-body is significantly stronger.  The inferred black-body
temperature in 2009 is only marginally higher than in 2006, but
the 2009 disc black-body normalisation is a factor $\sim30$ larger
than in 2006. The larger inferred inner disc radius in 2009 conflicts
with standard expectations for reduced disc truncation as the disc
luminosity increases.  However, the data can be reconciled with a
non-changing or even a smaller inner radius in 2009, if the intrinsic
disc temperature in 2006 is significantly lower than inferred from the
single disc blackbody fit and there is additional blackbody emission
from a hot inner ring, due to X-ray heating by the relatively stronger
power-law in 2006. 

\item  The PSDs in 2009 are band-limited, and can be interpreted in terms of two broad frequency components which peak around 0.4~Hz and 2~Hz respectively.  
\item The 2009 PSD shape and amplitude evolves with energy.  The
lowest-amplitude variability over all frequencies is seen in the
softest 0.3--0.6~keV band, followed by the 0.6--1.0~keV band. The
amplitude of the lowest-frequency component is largest in the 1.0--2.5 keV band before dropping slightly at higher energies, while the amplitude of the high frequency component increases with energy.
\item The covariance and rms spectra suggest that the energy-dependent PSD behaviour is linked to two effects.  Firstly, there is a fall-off in fractional rms (seen in all frequency ranges) at low-energies where the disc black-body dominates the emission.  Secondly, the variable power-law emission component becomes harder at higher temporal frequencies.  Neither of these effects are seen in the 2006 data.  In fact, in 2006 the opposite is seen at low energies, where the fractional rms increases in the energy range where the disc contributes to the spectrum.
\item Frequency-dependent lag measurements show that in general, the
harder photons lag softer ones.  Within the energy bands dominated by
the power-law (2.5--4~keV and 4--10~keV) and the disc (0.3--0.6~keV and
0.6--1~keV) the lags are relatively short and the phase-lags evolve
only smoothly and weakly (if at all) with frequency.  However, the
lags between power-law and disc bands are larger and show significant
structure in the form of `steps', with a significant increase in phase lag with frequency.  
\item Lag-energy spectra show similar behaviour to that reported by us
\citep{uttley2011} in other hard state observations of BHXRBs (including SWIFT~J1753.5--0127 in 2006).  The sharpest change in the lags is seen at low energies where the disc dominates the spectrum.  The low-energy shape of the lag spectrum depends on the frequency probed: the highest frequencies show evidence for a shift in the break in the lag spectrum to higher energies, together with an upturn in the lowest energy bins (similar to that seen in GX~339-4, \citealt{uttley2011}).
\item The coherence between the bands dominated by the power-law
remains high over a broad frequency range.  However, between the
power-law and disc-dominated bands, and between the two
disc-dominated bands, coherence drops significantly at frequencies
below 0.5~Hz (where the softest disc band shows the least coherence
with the other bands) as well as above 1~Hz (where both disc bands show low coherence with respect to the power-law bands).  Comparison of the covariance and rms spectra confirm that low-frequency variations of the softest photons, below 0.5 keV, are only weakly correlated with variations at higher energies.  In contrast, the 2006 data are relatively coherent at low frequencies.
\end{enumerate}
The spectral-timing behaviour of SWIFT~J1753.5--0127 in 2009 is complex
but we will attempt to interpret it here in terms of the physical picture
of mass-accretion rate fluctuations arising in and propagating through
the accretion disc, suggested by \citet{wilkinson2009} and
\citet{uttley2011} in order to explain the covariance and time-lag
behaviour of hard state BHXRBs.  Since we see similar lag behaviour in
the SWIFT~J1753.5--0127 2009 data, it is useful to
understand whether this model can be extended to explore the more
complex spectral-timing behaviour seen in this dataset.  In doing so,
we will focus primarily on explaining the differences between the spectral-timing properties of the 2009 data we present here and those from 2006 and observations of other BHXRBs reported in \citet{wilkinson2009} and \citet{uttley2011}.  Comprehensive testing of the propagating fluctuations model against these data is beyond the scope of this work, so here we simply sketch how the key observational results might be explained by the model.

\subsection{Implications from the propagating fluctuations model}

The key feature of the propagating fluctuations model used to explain
the soft X-ray behaviour of other hard state BHXRBs is that the disc
plays an important role in generating and carrying the fluctuations (at least
on fluctuation time-scales $>1$~s). 
In Subsection~\ref{subsec:emissivity} we assumed that the intrinsic
disc emissivity profile is the same in both observations ($\propto
R^{-3/4}$). Under this scenario, we showed that the disc emission in
2006 may be explained if the overall disc emission in 2006 is the result of
the extended emission with a `standard' emissivity profile, plus a
thermal emission component that is caused by significant heating of
the disc by the illuminating power-law photons. 

This additional inner-disc emission produces up to $\sim 75$ per cent of the disc
emission in the soft band (Fig.~\ref{fig:emissivity}). Due to the lower disc
temperature and the relatively hot inner ring, the disc emission in
2006 is much more centrally concentrated than in 2009. Furthermore,
the ultrasoft emission is significantly more extended across the disc
than the soft emission. 

We can use these differences to explain some
of the interesting spectral-timing behaviour in the data.  Note that
the arguments that follow apply independently of the actual inner disc
radius $r_{\rm in}$, since the emission profile in each band is set by
the disc temperature at $r_{\rm in}$ (which is set by the spectral fit
results)  and the {\it relative} change in radius.

First, let us consider the 2009 data and suppose that a fluctuation in
the disc arises at a relative small radius $R\sim1.5$~$r_{\rm in}$,
and modulates the disc emission as it propagates inwards.  Since only
$\sim$40~per~cent of the ultrasoft flux is contained within this
radius, the observed variability amplitude will be suppressed by a
factor $\sim2.5$ compared to the amplitude of the intrinsic
fluctuation.  In contrast, the soft band fluctuation will be
suppressed by only a factor $\sim1.5$, while the power-law emission,
if it originates within the disc truncation radius, will not be
suppressed at all.  This simple picture can explain the energy
dependence of the PSD amplitude in 2009 and the corresponding drop in
rms and covariance spectra at low energies.  In this scenario, the
same behaviour may not seen in other hard-state \textit{XMM-Newton}
observations of BHXRBs or in the 2006 data because of two effects.  Firstly, since the PSDs show lower characteristic frequencies in these other observations, the fluctuations in accretion rate may arise at a larger radius, so that a significantly larger fraction of the disc emission can be modulated.  Secondly, as can be seen in the 2006 emission profile, the combination of X-ray heating of an inner ring with a cooler disc at larger radii will lead to more concentrated emission in both disc bands, so that all or most of the flux will be modulated by a fluctuation arising at $R\sim1.5$~$r_{\rm in}$ or larger.  Thus we do not see suppression of the PSD amplitude at soft X-ray energies.  

\subsubsection{Coherence and frequency-dependent spectra}

We next consider the drop in coherence at low frequencies.  The natural explanation for this behaviour is that there is some component of accretion fluctuations in the disc which are unable to propagate very far inwards and therefore appear as variations in the extended disc emission (in the ultrasoft band and to a lesser extent the soft band) but do not produce correlated variations in the power-law emission. One possibility is that these variations are generated at larger radii than the spectrally-coherent fluctuations, so that they are viscously damped before they are able to propagate to the innermost part of the disc.  Again, we may see this effect in the 2009 data because of the strong intrinsic disc emission and extended emissivity profile.  Disc emission variations driven by power-law heating of the disc will be much more coherent with the power-law variations.  Furthermore, the lower-frequencies seen in the PSD of the 2006 data could imply that signals can effectively propagate from larger radii in the disc and thus modulate a significant fraction of the intrinsic disc emission coherently. 

The higher-frequency drop in coherence corresponds roughly to the
overlap between the low and high-frequency PSD components, and could
be produced by a similar effect to that described above.  For example,
if the higher-frequency signal corresponds to a second, even smaller
radius in the disc which generates propagating accretion fluctuations,
the signals generated at the same frequencies at larger radii may not
propagate effectively through the disc, again producing fluctuations that are not seen in the power-law component.  It is important to note however that the relatively small radii which we suggest as the origin of fluctuations (e.g. within 1.5~$r_{\rm in}$) also imply that the power-law emitting region must be compact. Otherwise, seed photons from the varying regions in the disc would upscatter to produce variations of the power-law that are coherent with the disc variations.  Similarly, we expect some component of disc emission to be produced by X-ray heating by the power-law and hence be coherent with the power-law variations.  These components are likely to exist at some level in the variable X-ray signal, but they must be relatively weak, at least on the time-scales where the coherence is low.

Some level of interaction between the disc photon variations and the
central power-law emitting region could potentially explain the observed hardening of the covariance and rms spectra at higher frequencies (and correspondingly, the peak in low-frequency power in the 1--2.5~keV energy range).  The amplitude of soft-photon variations is largest in the low-frequency peak of the PSD.  Compton cooling of the power-law emitting region by these soft photons could lead to a significant pivoting of the power-law in response to these variations, such that at low frequencies the fractional amplitude of power-law variations increases towards lower energies.  This effect then has the converse implication that at high frequencies the variable power-law appears to harden.

\subsubsection{Lags}

Finally, we consider the lag behaviour.  The relatively small lags and
smooth dependence on frequency seen between the two power-law bands and
also the two disc bands can be simply explained because here we are seeing
the same spectral components.  Fluctuations propagating through the
disc  will primarily modulate the ultrasoft emission first, but there
is significant overlap between soft and ultrasoft emission across the disc,
so we expect lags to be relatively small.  The same is true for the
power-law component -- lags should be small between energies where the power-law component
dominates.
The largest lags are then expected between the disc
and the power-law components, assuming that these are physically separated,
i.e. the power-law emission primarily originates within the disc
truncation radius.  

We must still explain the clear structure in the lag vs. frequency
dependence of the power-law relative to the disc-dominated bands, i.e.
the apparent `steps'.  Similar steps can be seen in other hard state
data from {\it RXTE}, but between power-law dominated energies
\citep{nowak1999}.  In the SWIFT~J1753.5--0127 2009 data, the effect
is particularly strong: the time lags actually increase with frequency
above $\sim 0.5$~Hz before dropping above $\sim1.5$~Hz. \citet{nowak2000} and
\citet{kotov2001} have suggested that steps in lag vs. frequency could
be related to the transition between two variability components, since
the steps seem to occur in between the `bumps' or broad Lorentzians in
the PSD.  If each component shows a different lag, one would expect a
stepped lag vs. frequency dependence consisting of discrete lags with relatively weak lag vs. frequency dependence where a single component dominates the variability, but with sharp steps occuring in the transition region between components.  This picture
may fit with the SWIFT~J1753.5--0127 2009 data, but the signal-to-noise is such that it is difficult to find a precise match between the step in the lag and the transition between the low and high-frequency components in the PSD (which seems to occur around 1~Hz). 
It is important to note however that the rise in the time lag seems to
coincide with the frequency range where the coherence peaks around
unity.  Incoherent signals which are caused by a mixture of
uncorrelated variable components will show lags which are a mixture of
the lags of each separate component.  Thus, the rise in lag could be
explained if the component causing the low-frequency drop in coherence
contributes a small lag, reducing the total observed lag, but at
higher frequencies of $\sim0.5$~Hz this component disappears from the
variable signal and the lag suddenly increases as a result.  One
possibility is that the incoherent fluctuations in the disc also modulate a small disc-corona component which is physically separate (and also spectrally distinct from) the inner power-law emitting region, which may be linked to a hot inner flow or the base of a jet.  The interband lags of such a component would then be relatively small, and would give way to the larger lags (due to propagation effects) once the coherent signal becomes dominant.  If the lags of the coherent signal themselves show a dependence on frequency, (e.g. a weak increase in phase lag with frequency as observed in other hard state BHXRBs) a complex pattern of rises and falls could be produced.

\section{Conclusions}

The soft X-ray coverage of the EPIC-pn camera onboard
\textit{XMM-Newton} has allowed us to carry out a comprehensive study of the spectral-timing properties of the BHXRB
SWIFT~J1753.5--0127 in the luminous 2009 hard state, and to perform a
comparison with the fainter 2006 hard state.

In our previous study of a sample of hard-state BHXRBs
\citep{uttley2011}, we showed that (i) soft (disc) X-ray variations
lead hard (power-law) variations 
up to frequencies of about 1 Hz, and (ii) at the
same low frequencies, the fractional variability in the disc is
larger than the variability observed in the power-law component.
Mass-accretion rate fluctuations that propagate through the disc
would produce variability on frequencies corresponding to the 
viscous time-scale where they are generated, and eventually reach a centrally-concentrated
source of power-law photons. This basic picture would both explain
the hard-to-soft lags and give a reasonable explanation for the excess
fractional variability seen in the disc as compared to the power-law.

The 2009 observation shows a breadth of previously unknown
observational features, which we summarise as follows.

Firstly,
the variability amplitudes in the PSD are strongly dependent on energy, with the
harder band varying more than the softer bands.
The explanation for this effect is seen in the fractional covariance 
spectrum, where the
bright disc black-body component that is seen in the spectrum now appears
suppressed in fractional-variability terms with respect to the
power-law, at all frequencies considered. This is opposite to what was
found in less-luminous hard states in \citet{uttley2011}, where the disc was more variable than the
power-law component. The higher luminosity (and thus, accretion rate) seen in this observation
could imply that the disc emissivity in the observed bands
is less centrally peaked than in
less-luminous hard states. In this case, a fluctuation that is
produced at an inner disc radius (e.g. within 1.5 $r_{\rm in}$) will not modulate the emission from outer
radii and this will effectively reduce the \emph{observed} fractional variability of
the disc as a whole. This is consistent with the shape of the
energy-dependent lags.

Secondly, we have found that the low frequency ($\lesssim 0.5$~Hz) disc variability below $< 0.6$~keV
appears partly uncorrelated with the rest of the emission produced by
both the disc and power-law components. An explanation for this could
exist if this incoherent accretion variability is produced at larger
radii than the coherent variability, affecting the softest emission
that is more extended while being unable to propagate to the innermost part of the disc and the power-law emitting region due to viscous damping.

Finally, the last feature to highlight is an up-turn in the frequency-resolved time lags
above $\sim 0.5$~Hz. This could be produced if, e.g. a relatively central secondary
extended corona that sandwiches the disc -- and thus reduces the observed
hard-to-soft lags caused by disc propagation variations -- becomes swamped by a coherent source of variations associated with the high-frequency peak in the PSD (perhaps an even smaller unstable radius in the disc).

By identifying a number of spectral-timing signatures with the accretion disc, this work suggests that a wealth of information regarding the
properties of accretion flows can be discovered by extending spectral-timing measurements to softer X-rays and using techniques that allow better spectral-resolution. By combining these powerful techniques with more detailed modelling, 
a comprehensive understanding of the variability may help to unravel
many of the mysteries underlying accretion physics and associated
phenomena such as the nature of state transitions and the production
of the jets seen in the hard states of accreting black holes.

\section*{Acknowledgements}
The authors would like to thank the referee for useful comments that
contributed to the clarity of the paper, J.~C. Houck for access to the useful intrinsic ISIS functions
that allowed us to perform a model convolution through the EPIC-pn
instrumental response in order to calculate the disc emission profile,
and M.~Hanke for useful advice.
The research leading to these results has received funding from the
European Community's Seventh Framework Programme (FP7/2007-2013) under
grant agreement number ITN 215212 ``Black Hole Universe''.

\label{lastpage}

\end{document}